# Versatile Variable Temperature and Magnetic Field Scanning Probe Microscope for Advanced Material Research

Jin-Oh Jung[a)], Seokhwan Choi[a)], Yeonghoon Lee, Jinwoo Kim, Donghyeon Son, Jhinhwan Lee[b)]

*Department of Physics, Korea Advanced Institute of Science and Technology, Daejeon 34141, Korea*

We have built a variable temperature scanning probe microscope (SPM) that covers 4.6 K - 180 K and up to 7 Tesla whose SPM head fits in a 52 mm bore magnet. It features a temperature-controlled sample stage thermally well isolated from the SPM body in good thermal contact with the liquid helium bath. It has a 7-sample-holder storage carousel at liquid helium temperature for systematic studies using multiple samples and field emission targets intended for spin-polarized spectroscopic-imaging scanning tunneling microscopy (STM) study on samples with various compositions and doping conditions. The system is equipped with a UHV sample preparation chamber and mounted on a two-stage vibration isolation system made of a heavy concrete block and a granite table on pneumatic vibration isolators. A quartz resonator (qPlus) based non-contact atomic force microscope (AFM) sensor is used for simultaneous STM/AFM operation for research on samples with highly insulating properties such as strongly underdoped cuprates and strongly correlated electron systems.

## I. INTRODUCTION

A stable and versatile SPM system with high magnetic field and wide variable temperature capability has been sought after in the fields of high-$T_c$ unconventional superconductors and strongly correlated electron systems for atomic scale verification of phase transitions in terms of structure, magnetism and superconductivity.[1-3] However, typical variable temperature SPM designs have required the scanner temperature close to the sample temperature, which made the high resolution quasiparticle interference measurement at sample temperature beyond 100 K over 12 hours through several days very difficult due to the strong scanner piezo droop rate nearly proportional to the scanner temperature. In order to overcome this issue, it has been required to thermally isolate the sample holder from the rest of the SPM head while using the small diameter Pan style SPM head suitable for high magnetic field experiments. Furthermore, recent researches on metal-insulator transition, Mott-insulating phase of cuprates, novel itinerant electronic states on the interface of insulators, and spintronic devices partly made of insulators has raised necessities of non-contact AFM technology simultaneously operable with conventional STM mode.[4]

## II. OVERALL SPECIFICATION GOALS AND ACHIEVEMENTS

Here we report a variable temperature medium-high field scanning probe microscope with following features:

(1) Two-stage vibration isolation system made of a 27-ton concrete air table and a 0.8-ton granite air table. In addition to the two stage isolation design which has become a standard, we used granite table with embedded tapped hole inserts rather than purely metallic (aluminum or stainless steel-based) structure for simplicity and rigidity.

(2) Temperature range of 4.6 K ~ 180 K. Its upper limit is about 130% of the $T_c$ (134 K) of $HgBa_2Ca_2Cu_3O_{8+\delta}$, the highest-$T_c$ superconductors under ambient pressure.

(3) Vertical magnetic field up to 7 T which creates about 10 vortices inside a *dI/dV* map area of (60 nm)$^2$ required for high resolution gap symmetry analysis using quasiparticle interference measurement and overcoming the coercive fields for many ferromagnetic spin-polarized tips. For upgradability to an extreme (> 20 T) magnetic field at a reasonable cost, we used the head design that fits in a small (52 mm) bore liquid-helium-cooled magnet.

(4) A 7-sample-holder storage carousel at 4.2 K temperature, suitable for 4 samples with different doping levels, 2 different field emission targets and one extra space for exchanging samples.

(5) UHV sample preparation chamber with an e-beam heater, an ion gun, a sample evaporator and a residual gas analyzer, for standard metallic sample surface preparation and general purpose leak detection.

(6) qPlus-sensor-based non-contact AFM system simultaneously operable with STM mode to achieve subatomic resolution imaging as well as to perform researches on possibly insulating samples and substrates.

---

a) J.-O. Jung and S. Choi contributed equally to this work.
b) Author to whom correspondence should be addressed: jhinhwan@kaist.ac.kr

## III. DESIGN AND CONSTRUCTION

Figure 1 shows the overall design of the low-vibration laboratory and the instrument. The SPM system is on a two-stage vibration isolation system made of a 27-ton concrete block (pink) with embedded SUS 304 rebar mesh structure and a 0.8-ton granite table (dark gray). The central bore of the granite table is covered with a thick Al plate for mounting a liquid helium dewar from below, using an electric winch with 600 kg capacity mounted on the ceiling. The helium dewar has belly capacity of 110 liters and the typical minimum liquid helium boil-off rate is about 8 liters per day with the retractable magnet current leads cooled by boiled off helium gas left connected to the magnet. As shown in Fig. 1(c), the two-stage vibration isolation system efficiently suppresses vibration modes above 50 Hz which contains most problematic modes to the STM measurement considering the rigidity of the STM head construction.

As shown in Fig. 2(a), on the granite table are a compact UHV sample preparation chamber with an ion pump, a turbo molecular pump, an e-beam sample heater, a residual gas analyzer, an ion gun, and full range pressure gauges. The two gauges are separated by a gate valve which can optionally shut off the cold UHV space during normal SPM measurement operation. Figure 2(c) shows the e-beam sample heater mounted on a retractable magnetic transporter. It has a cathode filament in close proximity to the sample holder and a K-type thermocouple in contact with the sample stud. Mounted on the four 10-pin UHV feedthroughs of the main chamber is an RF-filtered switch box of Fig. 2(b) for instrumentational signals such as thermometers, piezo scanner signals, walker signals, etc. On the three independent SMA feedthroughs are a bias voltage divider box and the current preamplifier boxes for tunneling current output and the qPlus sensor oscillation current output.

As shown in Fig. 3(a), the SPM head fits inside a 52 mm bore magnet while the 4 K flange is indium sealed and has 4.5-inch diameter. We directly mounted the 7-Tesla NbTi magnet on the vacuum can and routed a pair of $Nb_3Sn$ bus bars and the mating stainless steel retractable leads through the grooves of the indium seal flange and the cut-outs on the sides of the baffles of the main dip stick chamber.

As shown in Fig. 3(b), just below the 4 K flange inside the vacuum are a key hole structure with attached temperature sensor for contact-cooling of the sample before cleaving, a horizontally swinging bucket sample cleaver, and a storage carousel for seven sample holders. Each of the sample cleaver bucket and the storage carousel is driven with a room-temperature rotational feedthrough coupled through a stainless steel tube.

Figure 3(c) shows the SPM head mounted on the bottom end of a thick copper tube whose upper part is in good thermal contact with the 4K flange and the liquid helium bath. We also thermally isolated the sample holder from the SPM head. This asymmetric thermal design is crucial in maintaining the temperature of the SPM head and the scanner as low as possible to achieve a minimal temperature-dependent scanner piezo drooping while the sample stage is heated to its upper limit (~ 180 K). This feature is difficult to achieve by the suspended STM head design with Eddy current damping due to difficulty in achieving sufficiently low thermal resistance of the fine wire braids without introducing excessive mechanical vibration path. Fig. 3(d) shows the overall thermal design in a simplified diagram for clarity. The sample stage has a 38-AWG manganin wire heater with net resistance about 200 Ohms and a CX-SD type Cernox temperature sensor for closed loop control with a Lakeshore 340 controller. The sample stage is mounted on the walker body via a polished radiation shield [dark blue plate near the scanner in Fig. 3(d)] and vespel insulating washers [dark red in Fig. 3(d)]. Another temperature sensor is mounted on the walker body for monitoring the SPM body (i.e. the walker, the scanner and the tip) temperature.

Regarding the sample holder design, we typically fasten a flanged single crystal metal sample inside a screw-capped sample holder as shown in the Fig. 2(c). For cleavable samples, we typically glue one using silver epoxy on a flat faced sample holder as shown schematically in the Fig. 3(d).

In order to have optimal high voltage driving waveforms suitable for Pan style piezo walker operating at cryogenic temperatures, we first generated a smooth pulse made of four parabolic curve segments using a digital signal processor (DSP) and an analog-to-digital converter (ADC) as shown in Fig. 4(a)-4(c). Its monopolar waveform with adjustable amplitude (up to 10 $V_{pk}$) is then amplified by a gain of 30 using a high-power op-amp PA93 before being fed into the six independent push-pull switches made of pairs of power MOSFETs.[5] Highly asymmetrical accelerations of the inertial piezo motor are achieved with the fast sliding phase (six individually switched signals with < 1 μs time constant) and the carrying phase (a commonly driven parabolic S-waveform over about 1 ms).

The walker is Pan style made of a sapphire prism and six shear piezo stacks.[6] As shown in Figs. 3(c) and 4(d), attached at the bottom of the moving sapphire prism is a cylindrical tube used for differential capacitance position sensing. This metallic tube moves inside two split tubes with slightly larger identical diameters and with a small gap in between. The ratio of the capacitances between the inner tube and the two outer tubes is measured by driving the two outer tubes with 180° out-of-phase sinusoidal wave with LabVIEW-controlled amplitudes by proportional-integral-derivative (PID) feedback control so that the capacitively picked up signal amplitude at the inner tube is always maintained at zero by perfect cancellation of the two driving signals as shown in Fig. 4(d). The driving amplitude ratio thus determined is equivalent to the capacitance ratio of the two split tubes with the inner moving tube, which can be converted to the position of the walker. This nulling PID feedback scheme ensures that the walker position readout is less dependent on the parallel shifts between the inner tube and the two outer tubes due to temperature variation and the asymmetry of the Pan style SPM head.

The scanner and z-feedback controller is made of the Nanonis SPM controller Base-Package v4.5 and a customized electronics with high voltage amplifiers for five scanner signals and two (positive and negative monopolar) high voltage field-emission signals, as well as a sample bias summing and levelling electronics and the differential capacitance position sensor electronics. Due to the 4.8 K minimum temperature corresponding to thermal broadening about 0.4 mV, we use an electrometer-grade op-amp (OPA128LM) with a room-temperature 1GΩ feedback resistor which performs similarly well compared with commercially available current amplifiers in terms of frequency characteristics and noise level, which is, however, not included as the main topics of this work.

For future researches on materials with potentially insulating properties, such as metal-insulator transitions, doped Mott insulators and antiferromagnetic semiconductors, as well as sub-atomic bond imaging of organic molecules,[4] we added a customized qPlus sensor to the SPM scanner as shown in Figs. 5(a)-(c). We used a qPlus resonator with nominal resonance frequency of 51 kHz from Scienta-Omicron (B803037) and a separate slab piezo for mechanically driving the whole resonator for minimal electronic cross-talk between the driving and the output oscillation signals. We used a 51 kHz sensor with built-in tip line in the quartz cantilever instead of a 32 kHz sensor that can be found in quartz watches to achieve better Q factor by eliminating possible loss in the externally bonded tip line as well as to achieve slightly higher scanning speed without loss in the sensitivity of the sensor. As shown in Fig. 5(d), we could achieve $f_{res}$=46,463Hz and Q factor of 4,028 at room temperature with a tip attached, slightly better than those reported in literature.[7]

## IV. PERFORMANCE

### A. Temperature Dependence Characteristic

With the current design of the polished radiation shield and vespel insulating washers, the temperature raise ratio is maintained at $\frac{T_{sample}-T_{keyhole}}{T_{head}-T_{keyhole}} \approx 10$ which results in $T_{head} \approx 23$ K when $T_{sample} \approx 180$ K and $T_{keyhole} \approx T_{LHe} = 4.2$ K as shown in Fig. 6(a). The insulating vespel washer dimensions are determined in such a way that their long length and thin cross-section do not induce mechanical weakness that is adverse to the SPM operation. As shown in Fig. 6(b), the helium boil-off rate at the maximum temperature of ~180 K is suitable for medium size *dI/dV* map that takes up to two days, while at the minimum temperature of ~5 K a high-resolution map taking up to 8 days is possible with the current dewar capacity. The sample temperature as a function of time after the feedback is turned on is plotted in Fig. 6(c). It takes about 1.5 - 2 hours before the temperature is stabilized to within ±1 mK to the target value.

After the temperature is stabilized, the drift rate is dominated by the piezoelectric scanner drooping as a result of large scanner voltage changes in preparation of the map. The droop rate can be modelled by $v_x(t) \approx \frac{\eta(T)\Delta x(t_0)}{t}$ where $\eta(T)$ is the unit-less piezo scanner droop coefficient and is an increasing function of scanner temperature, while $\Delta x(t_0)$ is the nominal displacement value (or the displacement measured at time $t_0 \sim 0.1$s after the stepwise change of the piezo voltage). As shown in Fig. 6(d), we have plotted the average values of $\eta(T)$ at three representative sample temperatures. We performed the fitting to $\frac{1}{v_x(t)} = \frac{t}{\eta(T)\Delta x(t_0)}$ of the apparent image drift rates measured with the first four successive atomic resolution scans immediately after 100 nm ($= \Delta x(t_0)$) step-wise movements in four (±X and ±Y) directions. Since $\eta(T)$ is a steeply increasing function of scanner temperature, it is clear that keeping the scanner temperature as low as possible is crucial in reducing the overall drift rate during a high resolution *dI/dV* map taking multiple days.

### B. STM/STS Experimental Results

As a first demonstration of the quality of the STM and STS data, we performed STM topography and spectroscopic imaging measurement at 4.6 K on underdoped cuprate superconductor $Bi_2Sr_2CuO_{6+\delta}$ (Bi2201). The large area topograph and its zoomed-in image [Fig. 7(a)] show high resolution typical of topographs in this STM. The Fourier-transformed 512-pixel 70-nm *dI/dV* image taken at 5 meV [Fig. 7(b)] shows quasiparticle interference peaks according to the octet model.[8-10]

As a part of the development of the spin-polarized STM technique, we have also performed spin-polarized STM topography measurement on $Fe_{1+y}Te$ over -7 T ~ 7 T using an Fe cluster tip. The spin polarization of the Fe cluster tip made of excess Fe atoms collected on the surface is reversed at -7 T and 7 T magnetic fields and the difference [Fig. 7(e)] of their topographs [Figs. 7(c)-7(d)] shows the spin-texture of the $Fe_{1+y}Te$ surface. This technique is used to confirm the sharpness and the spin-polarization of the tip before exchanging the sample holders via the sample storage carousel and performing the spin-polarized measurements on the target sample of scientific interest.

As an example of magnetic phase transition in Fe-based superconductor triggered by spin-polarized tunneling current, we have also taken variable temperature spin-polarized STM topography on $Sr_2VO_3FeAs$ over 4.6 K ~ 180 K using an antiferromagnetic Cr cluster tip.[11-13] It shows $C_4$ magnetic domains and phase domain walls [Fig. 7(f)] induced on the frustrated heterostructure Fe-based superconductor by spin-polarized current injection at 4.6 K. Beyond the Fe magnetic ordering temperature near 50 K (Ref. 14), the quasi-$C_2$ surface reconstruction is visible [Fig. 7(g)]. At even higher temperature beyond 150 K, no surface reconstruction is observed [Fig. 7(h)]. This magnetic phase transition below 50 K is shown to be coupled with the local superconductivity possibly via the spin-fluctuation-based theory of iron-based superconductivity.[13]

## V. CONCLUSION

We have developed and demonstrated the performance of a scanning probe microscope with novel thermal and mechanical designs for variable temperature and high magnetic field measurements of spin-polarized STM images and quasiparticle interference maps that require multiple days of data acquisition. The key features of the newly developed SPM system are as follows; (1) Two-stage vibration isolation system made of a 27-ton concrete air block and a 0.8-ton granite air table efficiently suppressing vibration modes above 50 Hz. (2) Variable temperature range of 4.6 K ~ 180 K with 1 mK stability, where the helium boil off rate at the 180 K is suitable for medium size $dI/dV$ map that takes up to two days, while at the 4.6 K a high-resolution map taking up to 8 days is possible. (3) Vertical magnetic field up to 7 T. (4) A 7-sample-holder storage carousel at 4.2 K for systematic and comparative sample studies. (5) UHV sample preparation chamber with an e-beam heater, an ion gun, a sample evaporator and a residual gas analyzer, for standard metallic sample surface preparation and leak detection. (6) Non-contact AFM mode simultaneously operable with STM mode. These features together may contribute to novel atomic scale verification of phase transitions in terms of structure, magnetism and superconductivity in the high-$T_c$ and unconventional superconductors and strongly correlated electron systems.

## ACKNOWLEDGEMENTS

We acknowledge contributions made by H. -S. Suh from Samsung Advanced Institute of Technology, by S. -M. Shin from Center for Applied Life Science in Hanbat National University, and by W. -J. Jang and Y. K. Semertzidis from Center for Axion and Precision Physics of Institute of Basic Sciences for the supports in the construction of the SPM. We also acknowledge contributions from the KAIST Institute, the department of physics, and the department of Energy, Environment, Water and Sustainability (EEWS) of KAIST for the support in the construction of the low vibration facility. Bi2201 samples are kindly provided by D. -J. Song and H. Eisaki from Advanced Institute of Science and Technology in Japan. Sr$_2$VO$_3$FeAs samples are kindly provided by J. M. Ok and J. S. Kim from department of physics in POSTECH, Korea. Most of the custom vacuum parts for the SPM are made by I.T.S., M&S and IdeaTech in Korea. The qPlus sensors are kindly supplied by Scienta-Omicron. We also thank J. C. Davis for the basic design of the SPM and F. J. Giessibl for the basic design of the qPlus sensor. This work was supported by National Research Foundation (NRF) through the Pioneer Research Center Program (No. NRF-2013M3C1A3064455), the Basic Science Research Program (No. NRF-2017R1D1A1B01016186) and the Brain Korea 21 Plus Program. It is also supported by IBS-R017-D1, Korea Research Institute of Standards and Science through the Metrology Research Center Program (No. 2015-15011069) and by the Samsung Advanced Institute of Technology (SAIT).

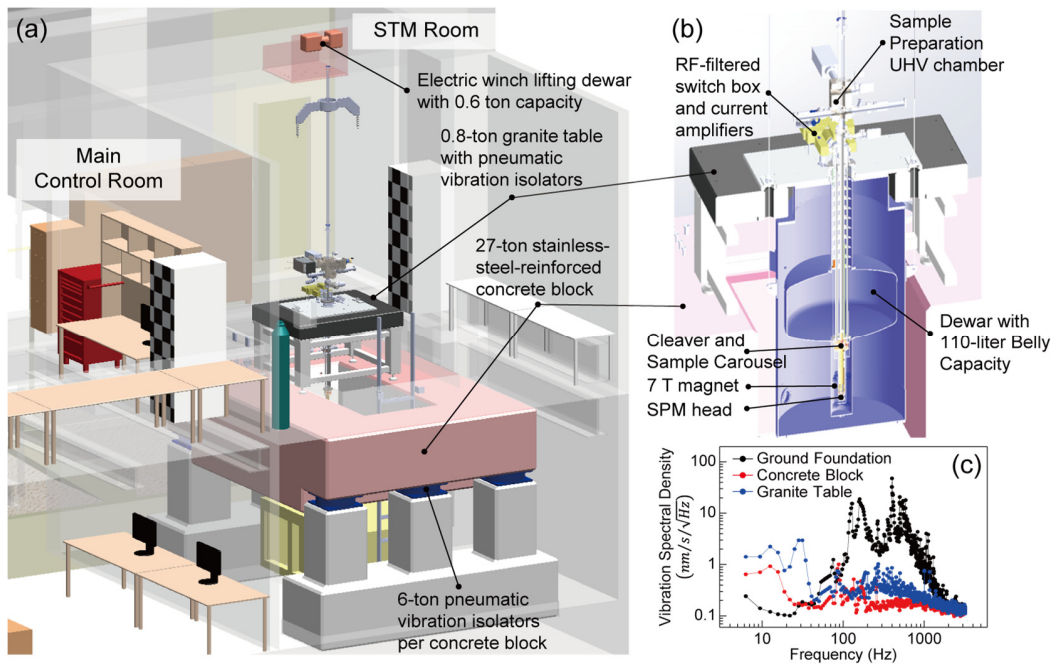

FIG. 1. (a) Overall view of the low-vibration laboratory. (b) Cross-section of the variable-temperature SPM. The SPM is mounted on a two-stage vibration isolation system made of a 27-ton stainless-steel-reinforced concrete block (pink) and a 0.8-ton granite table (dark gray) floated on a regulated 4-bar compressed air. The helium dewar (purple) has belly capacity of 110 liters and a typical minimum boil-off rate about 8 liters per day with magnet leads connected to the magnet. (c) Typical vibration spectra of the foundation, the concrete block and the granite table.

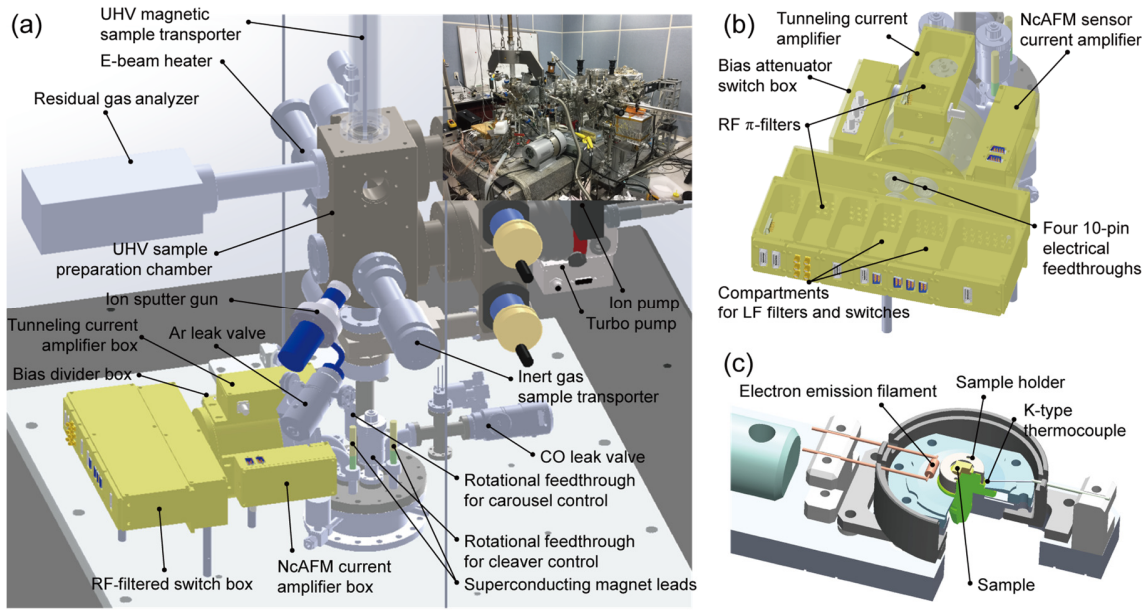

FIG. 2. (a) The UHV sample preparation chamber is equipped with a vertical magnetic sample transporter, an ion pump, a turbo molecular pump, an e-beam sample heater, a residual gas analyzer, an ion gun and two full range pressure gauges decoupled by a gate valve. The gate valve shuts off the vacuum space inside the dewar during normal SPM measurement operation. The inset shows the picture of the overall SPM system. (b) Also directly mounted on the UHV feedthroughs are a filtered switch box for instrumentational signals, a bias voltage divider box, three current preamplifier boxes for the tunneling current, the non-contact AFM sensor oscillation signal and differential capacitive walker position sensor signal. (c) The close up view of the e-beam sample heater with a tungsten filament cathode and a thermocouple in thermal contact with the sample holder. The sample holder shown is with an open screw cap fastening a flanged single crystal sample.

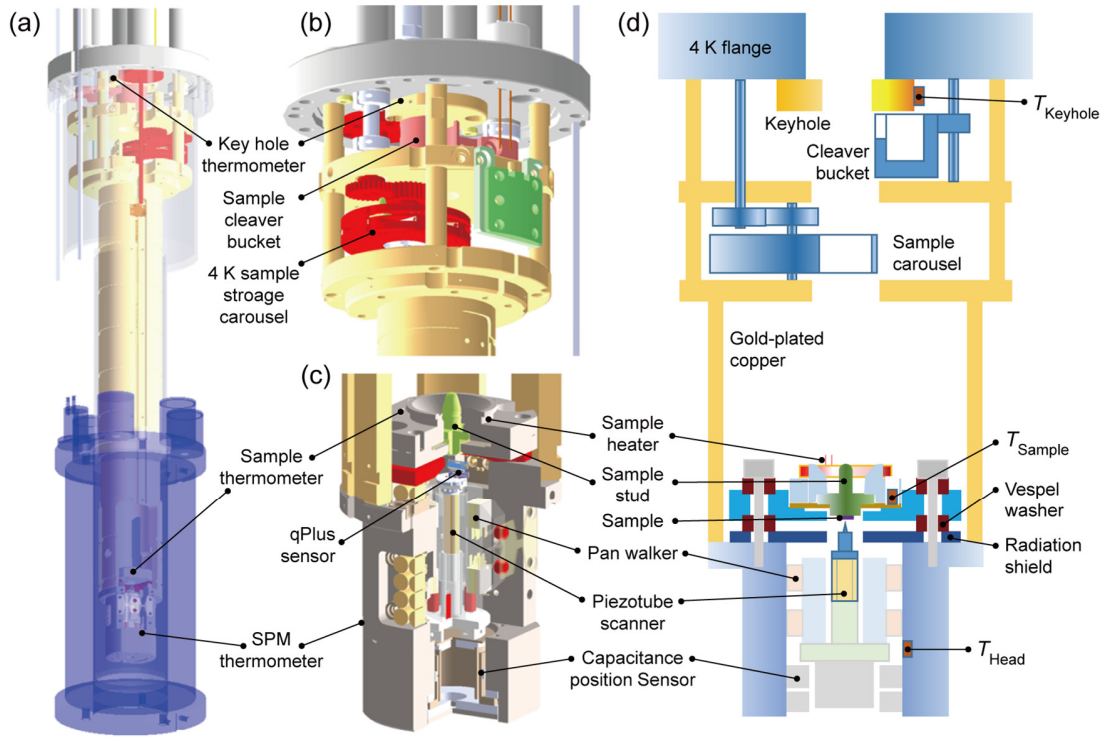

FIG. 3. (a) The main cryogenic part of the SPM with the thermometer locations indicated. The purple part at the bottom is a persistence mode 7-Tesla NbTi superconducting magnet. (b) A close-up view of the upper cryogenic part containing a swinging bucket sample cleaver (pink), a seven sample storage carousel (red) and a key hole with a thermometer (yellow). (c) A close up view of the SPM head attached at the lower end of the cryogenic part. On the top of the head is the sample mount with a manganin wire heater and a temperature sensor. A Pan-style walker is used with a differential capacitive walker position sensor mounted at the bottom of the head. Mounted at the upper end of the scanner is a receptacle for a non-contact AFM/STM (qPlus) sensor. (d) Thermal schematic diagram for the cryogenic part of the variable-temperature SPM. The sample temperature sensor is glued to the annular BeCu plate spring (dark orange) that clamps the sample holder to the sample mount with a significant force upon insertion. Shown in green is a flat-faced sample holder where a cleavable sample is glued with silver epoxy.

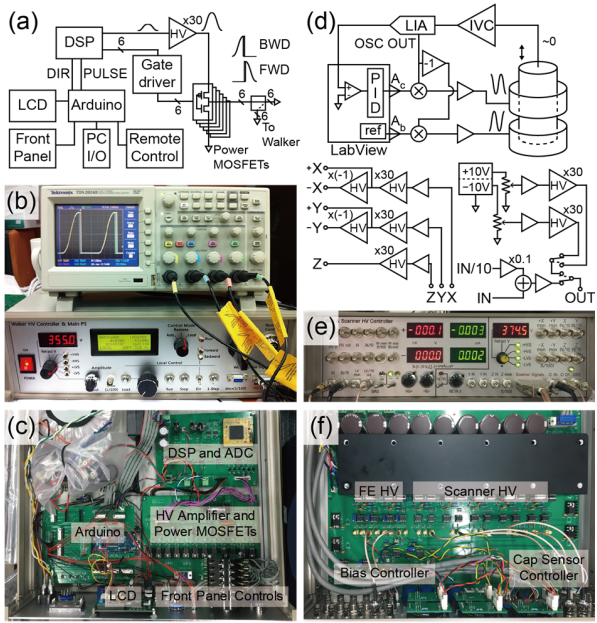

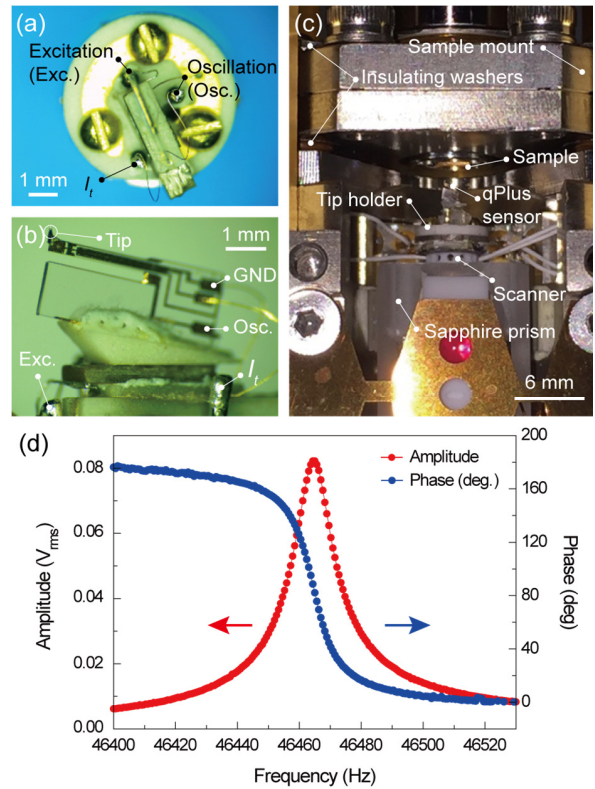

FIG. 4. (a) The schematic diagram of the walker control circuit with DSP-based S-waveform generator and six independently controlled push-pull MOSFET switches. (b) The high voltage output waveforms from the walker controller and its front control panel. (c) Printed circuit boards (PCBs) and interconnects in the walker controller box. (d) The schematic diagram of (top) the capacitive walker position sensor monitoring circuit with the calibrated output, and (bottom) the scanner high voltage (HV) controller, the field-emission HV sources and the bias voltage control circuit. (e) The front control panel and (f) its PCBs and interconnects inside. Nanonis Base-Package v4.5 is connected to these custom electronics.

FIG. 5. (a)-(c) qPlus sensor (51 kHz model B803037, Scienta-Omicron) mounted on the STM scanner. The gold-patterned alumina disc in (a) has diameter of 6 mm. (d) Resonance spectrum of the qPlus sensor mounted on the SPM scanner with the piezo driving voltage amplitude of 5 mV$_{pk}$.

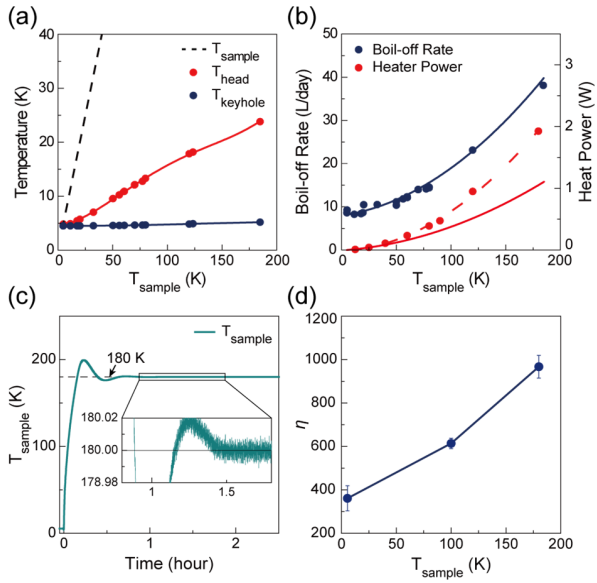

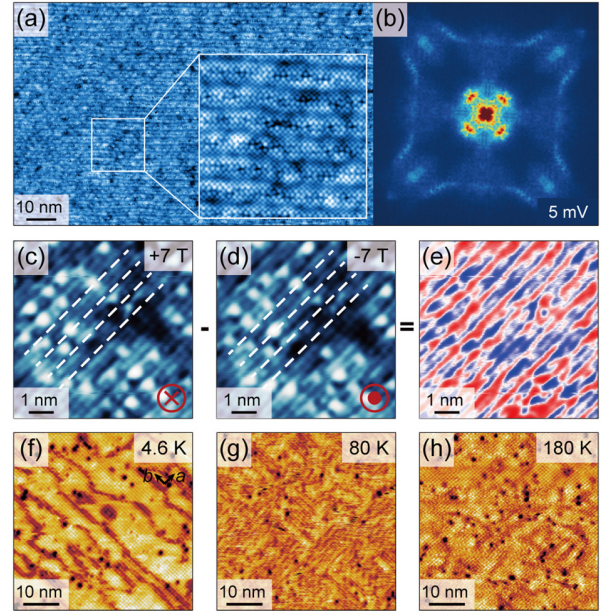

FIG. 6. Sample-temperature-dependence of (a) the SPM head temperature and (b) the helium boil-off rate and the required heater power. The dashed red curve in (b) is the total power supplied by the temperature controller and the solid red curve is the power dissipated solely below the liquid helium level used to heat the SPM head. The discrepancy corresponds to the heat dissipated in the cable. (c) Sample temperature as a function of time after the feedback is turned on. It takes about 1.5-2 hours before the temperature is stabilized to within ±1 mK if the random sensor noise is averaged out. (d) The piezo scanner droop coefficient $\eta$ plotted as a function of sample temperature. The error bars indicate the standard error of fitting, further reduced by averaging for step-wise motions in four different directions.

FIG. 7. (a) Normal STM topographic image on the surface of cuprate superconductor Bi2201 taken at 4.6 K for 2 hour with 1980×1080 resolution. The inset on the right is the actual digitally zoomed-in image of the area in the white square on the left. (b) Quasiparticle interference data (Z map at 5 meV) taken on the same surface, based on a (512 pixel)$^2$ (60 nm)$^2$ $dI/dV$ data with 201 energy layers. (c)-(d) SP-STM topographs of Fe$_{1+y}$Te with 256×256 resolution taken at 7 T and -7 T for 5 minutes each and (e) their difference image showing surface spin signals. (f)-(h) Variable temperature topographs with 256×256 resolution taken over 4.6 K ~ 180 K for 10 minutes each with antiferromagnetic Cr-tip on magnetically frustrated heterostructure iron-based superconductor Sr$_2$VO$_3$FeAs.[12,13]